\documentclass[12pt]{iopart}
\usepackage{amsfonts}
\usepackage{color}
\expandafter\let\csname equation*\endcsname\relax
\expandafter\let\csname endequation*\endcsname\relax
\usepackage{amsmath}
\usepackage{amssymb}
\usepackage{graphicx}
\newcommand{\be}{\begin{equation}}
\newcommand{\ee}{\end{equation}}
\def\ba{\begin{aligned}}
\def\ea{\end{aligned}}

\newcommand{\bea}{\begin{eqnarray}}
\newcommand{\eea}{\end{eqnarray}}

\begin{document}

\title{Casimir torque between two inhomogeneous semi-transparent concentric cylinders}

\author{M. Shahzamani, M. Amini and M. Soltani}

\address{Department of Physics, University of  Isfahan,
Isfahan, Iran } 

\begin{abstract}
Motivated by the problem of Casimir energy, we investigate the idea of using  inhomogeneity of surfaces instead of their corrugation, which leads to Casimir interaction between two  inhomogeneous semi-transparent concentric cylinders. Using the  multiple scattering method, we study the Casimir energy and torque between the cylinders with different potentials subjected to Dirichlet boundary conditions, both in weak and strong coupling regimes. We also extend our formalism to the case of two inhomogeneous dielectrics in a weak coupling regime.  
\end{abstract}

\maketitle

\section{Introduction}
Since the  pioneering work in 1948 \cite{Casimir1}, in which Casimir  showed that
quantum fluctuations of the electromagnetic fields can
generate measurable long-ranged forces between conducting plates,
a large and still-growing body of work has  generalized the idea to various geometrical configurations~\cite{Milton1,Milton2}.
Although from  a theoretical point of view, the configuration of  parallel plates is more accessible, it is much more preferable
both in experiment and application to consider  curved surfaces such as a lens, sphere or cylinder.
For these configurations one may consider the angular analogue of the Casimir force, the Casimir torque, which is its other important aspect and plays a major role in the recently developed nano-motors idea that was first proposed in~\cite{Barash}.
The torque between two
corrugated conducting plates   and also for concentric corrugated conducting cylinders
for the scalar field  was calculated  after that \cite{Lombardo,Cavero,Rodrigues}.

People used different methods to calculate the Casimir effect.
For two conducting plates, Casimir derived the force directly based on  a  change in
the (infinite) zero-point electromagnetic field energy~\cite{Casimir1}.
After that, Lifshitz \cite{Lifshitz1} considered  a more general approach based on electromagnetic
fluctuations in thermal equilibrium,  for  two dielectric half-spaces separated by a vacuum.
Then, the path integral methods  introduced by Bordag et al in \cite{Bordag}, used for two parallel perfect metal plates.
The quantum extension of this method was developed  by Golestanian and Kardar later \cite{Golestanian1,Golestanian2} which was then applied to the case of two plates with roughness \cite{Golestanian3}. The interpretation  of  the Casimir effect in terms of the radiation pressure associated with zero-point energy was another method for these kind of studies which is reviewed in \cite{Milonni} and used by Matloob \cite{Matloob}. Finally, there is the formalism of multiple scattering method proposed by Milton \cite{Milton} and Rahi \cite{Rahi} which  it is possible to use for different geometries; we used it in our study here.


The idea of calculating the  Casimir interaction between two  cylinder configurations has been considered in various papers (e.g.  \cite{Cavero,Mazzitelli1,Dalvit,Mazzitelli2}). 
In addition,   the  strategy  of computing the Casimir torque for  corrugated cylinders is discussed by Cavero-Pleaez in \cite{Cavero}. Generalizing this approach,  in this paper, in order to compute the Casimir torque between two concentric cylinders, we try to investigate an idea based on using inhomogeneous surfaces instead of corrugated ones which can be one of the interesting aspects of Casimir phenomena \cite{Bao}. 


In this paper we calculate  the Casimir energy and torque  between two inhomogeneous semi-transparent concentric cylinders using the multiple scattering method \cite{Milton, Rahi} with a massless scalar field in the presence of the Dirichlet boundary condition. We study the quantum fluctuation induced by Casimir energy  to find  the equilibrium points and maximum torque between the above geometries  both in weak and strong coupling limits. In the weak coupling limit we approximate the Casimir torque  analytically but in the strong coupling limit we use a numerical method to obtain energy and torque.
We also address the problem of two dielectric cylinders in the weak coupling limit.  

\section{Multiple scattering formalism of Casimir energy}

In this section we will introduce the multiple scattering formalism of Casimir energy which is discussed in detail in  \cite{Milton, Rahi}.  
It can be shown  that for an electromagnetic (EM) field in  the presence of axial symmetry, the field can be divided into TM
and TE modes which refer to the Dirichlet and Neumann boundary conditions respectively.  
Therefore, any EM field can be described by  two independent scalar fields \cite{Golestanian3,Soltani}, which we consider here.

For an equilibrium system described by a scalar field $\varphi$, 
the Casimir energy can be written as:
\begin{equation}
   \varepsilon _0  =  - \frac{{\hbar c}}{{2\pi }}\int\limits_0^\infty  {d\bar{\kappa} \;\log \;Z(\bar{\kappa})} 
\end{equation}
in which $Z(\bar{\kappa})$ is the partition function of the system defined as
\begin{equation}
   Z(\bar{\kappa})=\int{D\varphi \,}D{{\varphi }^{*}}\exp [-\beta \int{dx\,}{{\varphi }^{*}}({{\bar{\kappa}}^{2}}-{{\nabla }^{2}})\varphi +{{\varphi }^{*}}V\varphi ],
\end{equation}
where we used the wick rotation notation in which $\bar{\kappa}=i\omega$.

In order to study the Casimir force in more general geometries, we use the multiple scattering approach, in which 
the scattering amplitudes and translation matrices are combined within a simple algorithm that allows  efficient numerical and analytical calculations of  the Casimir force and torque for a wide variety of geometries, materials, and external conditions.
In this approach the Casimir energy can be expressed with
\begin{equation}
   \varepsilon =\frac{\hbar c}{2\pi }\int\limits_{0}^{\infty }{d\bar{\kappa} \ \log \ \det (MM_{\infty }^{-1})}
\label{Cas-en1}
\end{equation}
where $M_{\infty }^{-1}$ is a block diagonal matrix diag 
$\left( {F_1 \;\;\;F_2 \quad F_3 \quad  \cdots \quad \quad } \right)$
and objects do not interact  at infinite separation.
In the case where we have  interaction between two objects, equation (\ref{Cas-en1}) can be expressed in terms of the $F$ and $X$ sub-matrices,
\begin{equation}
   \varepsilon _0  = \frac{{\hbar c}}{{2\pi }}\int\limits_0^\infty  {d\bar{\kappa} \;\log \;\det ({\rm I} - F_a X^{ab} F_b X^{ba} )}, 
\label{eq4}
\end{equation}
in which scattering and translation matrices $F$ and $X$ will be introduced  below.
Since the Casimir force between macroscopic objects depends on their shapes and orientations, the Green's function representations   in various coordinate systems are crucial to this formalism.
So, we can write the equation of motion for free scalar field as:
\begin{equation}
   (-{{\nabla }^{2}}+{{\bar{\kappa}}^{2}})\varphi (\bar{\kappa} ,\operatorname{x},{x}')=0,
\label{eq5}
\end{equation}
and  Green's function as: 
\begin{equation}
   (-{{\nabla }^{2}}+{{\bar{\kappa}}^{2}}){{G}_{0}}(\bar{\kappa},\operatorname{x},{x}')=\delta (\operatorname{x}-{x}').
\label{g0}
\end{equation}
The  Green's function, in general, can be expressed as:
\begin{equation}
   G_0  = \sum\limits_\alpha  {C_\alpha  (\bar{\kappa} )} \;\left\{ \begin{array}{l}
 \varphi _\alpha ^{out} (\bar{\kappa} ,\xi _1 ,\xi _2 ,\xi _3 )\varphi _\alpha ^{reg*} (\bar{\kappa} ,\xi _1 ,\xi _2 ,\xi _3 )\;\;\;\;\;\;\;\;\xi _1 (x) \ge \xi _1 ^\prime  (x') \\ 
 \varphi _\alpha ^{reg} (\bar{\kappa} ,\xi _1 ,\xi _2 ,\xi _3 )\varphi _\alpha ^{in*} (\bar{\kappa} ,\xi _1 ,\xi _2 ,\xi _3 )\;\;\;\;\;\;\;\;\;\xi _1 (x) \le \xi _1 ^\prime  (x')\; \\ 
 \end{array} \right.
\label{green}
\end{equation}
in which $\xi$ denotes the coordinate space e.g. for a  cylinder $\xi=\rho$.
In this representation, we employed the ``regular'' solution which satisfies eq.~(\ref{eq5}) and is defined at the origin, while the ``outgoing" solution  is used for other positions. Also the $C_\alpha  (\bar{\kappa})$ is some normalization factor.
In the presence of the potential of object $i$, the scalar field and Green's function satisfy:
\begin{equation}
   (-{{\nabla }^{2}}+{{\bar{\kappa}}^{2}}+{{V}_{i}})\varphi (\bar{\kappa} ,\operatorname{x},{x}')=0
\end{equation}
and
\begin{equation}
   (-{{\nabla }^{2}}+{{\bar{\kappa}}^{2}}+{{V}_{i}})G(\bar{\kappa} ,\operatorname{x},{x}')=\delta (\operatorname{x}-{x}')
\end{equation}
The Lippmann-Schwinger  general solution of this equation reads as follows \cite{LS}: 
\begin{equation}
\left| \varphi  \right\rangle  = \left| {\varphi _0 } \right\rangle  - G_0 T\left| {\varphi _0 } \right\rangle, 
\end{equation}
where the T-operator is defined as:
\begin{equation}
T=V\frac{1}{1+{{G}_{0}}V}=V-V{{G}_{0}}V+V{{G}_{0}}V{{G}_{0}}V+....
\label{eq11}
\end{equation}
Now depending on geometry, we require the scattering amplitudes for different conditions, so we can consider two different types of scattering.
First we start with the outside scattering and we will use the homogeneous solution that is the regular wave function.
Suppose that the incident wave comes from the outside, so one can write the outside scattering as:
\bea
&&\left| \varphi (\bar{\kappa}) \right\rangle =\left| \varphi _{\alpha }^{reg}(\bar{\kappa}) \right\rangle +  \\ 
&&\sum\limits_{\beta }{\left| \varphi _{\alpha }^{out}( \bar{\kappa}) \right\rangle }\times (-1){{C}_{\beta }}(\bar{\kappa} )\underbrace{\left\langle \left. \varphi _{\alpha }^{reg}(\bar{\kappa} ) \right| \right.\left. T( \bar{\kappa}) \right|\left. \varphi _{\beta }^{reg}(\bar{\kappa} ) \right\rangle }_{F_{\beta ,\alpha }^{ee}},\nonumber
\eea
and the inside scattering as:
\bea
&&\left| \varphi ( \bar{\kappa} ) \right\rangle =\left| \varphi _{\alpha }^{reg}(\bar{\kappa}) \right\rangle +\\ 
&&\sum\limits_{\beta }{\left| \varphi _{\beta }^{reg}(\bar{\kappa}  ) \right\rangle }\times (-1){{C}_{\beta }}(\bar{\kappa} )\underbrace{\left\langle \left. {{\varphi }_{\beta }}^{in}(\bar{\kappa}  ) \right| \right.\left. T(\bar{\kappa} ) \right|\left. \varphi _{\alpha }^{reg}(\bar{\kappa}  ) \right\rangle }_{F_{\beta ,\alpha }^{ie}}, \nonumber
\eea
where $F_{\beta ,\alpha }^{ee}$ and $F_{\beta ,\alpha }^{ie}$ define the exterior/exterior and interior/exterior scattering amplitudes,
respectively.
The second type of geometry is the case in which  one object is inside the other and the homogeneous solution is  the outgoing wave function.
Suppose that the incident wave comes from inside so we can write the inside scattering as:
\bea
&&\left| \varphi (\bar{\kappa} ) \right\rangle =\left| \varphi _{\alpha }^{out}(\bar{\kappa}  ) \right\rangle +\\
&&\sum\limits_{\beta }{\left| \varphi _{\beta }^{reg}(\bar{\kappa}  ) \right\rangle }\times (-1){{C}_{\beta }}(\bar{\kappa} )\underbrace{\left\langle \left. \varphi _{\beta }^{in}(\bar{\kappa} ) \right| \right.\left. T(\bar{\kappa}  ) \right|\left. \varphi _{\alpha }^{out}(\bar{\kappa}  ) \right\rangle }_{F_{\beta ,\alpha }^{ii}},\nonumber
\eea
and the  outside scattering as:
\bea
&&\left| \varphi (\bar{\kappa} ) \right\rangle =\left| \varphi _{\alpha }^{out}(\bar{\kappa}  ) \right\rangle +\\
&&\sum\limits_{\beta }{\left| \varphi _{\beta }^{out}(\bar{\kappa} ) \right\rangle }\times (-1){{C}_{\beta }}(\bar{\kappa}  )\underbrace{\left\langle \left. \varphi _{\beta }^{reg}(\bar{\kappa} ) \right| \right.\left. T(\bar{\kappa} ) \right|\left. \varphi _{\alpha }^{out}(\bar{\kappa}  ) \right\rangle }_{F_{\beta ,\alpha }^{ei}}.\nonumber
\eea
It is now convenient to assemble the scattering amplitudes for inside and outside into a single matrix:
\begin{equation}
F(\kappa ) = \left( 
{\begin{array}{*{20}c}
      {F^{ee} (\bar{\kappa} )} & {F^{ei} (\bar{\kappa} )}  \\
      {F^{ie} (\bar{\kappa} )} & {F^{ii} (\bar{\kappa} )}  
\end{array}}
 \right).
\end{equation}
So if one has the scattering matrix of object $i$ in the coordinate system  $i$ and object $j$ in the coordinate system  $j$, then it is possible to expand the wave function in  coordinate system $i$ with respect to the system $j$ :
\begin{equation}
   \varphi _{\alpha }^{out}(\bar{\kappa} ,{{\operatorname{x}}_{i}})=\sum\limits_{\beta }{u_{\beta \alpha }^{ji}(\bar{\kappa} ,{{\operatorname{x}}_{ji}})\,}\varphi _{\beta }^{reg}(\bar{\kappa} ,{{\operatorname{x}}_{j}}),
\label{phi_o}
\end{equation}
\begin{equation}
   \varphi _{\alpha }^{reg}(\bar{\kappa} ,{{\operatorname{x}}_{i}})=\sum\limits_{\beta }{\nu _{\beta \alpha }^{ji}(\bar{\kappa} ,{{\operatorname{x}}_{ji}})\,}\varphi _{\beta }^{reg}(\bar{\kappa} ,{{\operatorname{x}}_{j}}).
\label{phi_r}
\end{equation}
If we combine  equations (\ref{phi_o}) and (\ref{phi_r}) with equation (\ref{green}), the free Green's function can be written as:
\begin{equation}
   G_0 (\bar{\kappa},{\mathop{\rm x}\nolimits} ,x') = \sum\limits_{\alpha ,\beta } {C_\beta  (\bar{\kappa} )\left\{ \begin{array}{l}
	    \varphi _\alpha ^{reg} (\bar{\kappa} ,{\mathop{\rm x}\nolimits} _i )u_{\alpha ,\beta }^{ji} \varphi _\beta ^{reg*} (\bar{\kappa} ,{\mathop{\rm x}\nolimits} _j ^\prime  )\;\;\quad \quad \quad  \text{if $i$ and $j$ are outside each other} \\ 
	    \varphi _\alpha ^{reg} (\bar{\kappa} ,{\mathop{\rm x}\nolimits} _i )\nu _{\alpha ,\beta }^{ji} \varphi _\beta ^{in*} (\bar{\kappa} ,{\mathop{\rm x}\nolimits} _j ^\prime  )\quad \quad \quad \quad  \text{if $i$ is inside $j$} \quad  \\ 
	    \varphi _\alpha ^{out} (\bar{\kappa} ,{\mathop{\rm x}\nolimits} _i )w_{\alpha ,\beta }^{ji} \varphi _\beta ^{reg*} (\bar{\kappa} ,{\mathop{\rm x}\nolimits} _j ^\prime  )\;\;\quad \quad \quad\text{if $j$ is inside $i$}, \quad  \\ 
 \end{array} \right.} 
\end{equation}
therefore, we can define the translation matrix with respect to the coefficients of expansion in a general situation like:
\begin{equation}
   X^{ij} (\bar{\kappa} ) = \left( {\begin{array}{*{20}c}
	 { - u^{ji} (\bar{\kappa} )} & { - \upsilon ^{ij} (\bar{\kappa} )}  \\
  { - w^{ji} (\bar{\kappa} )} & 0  
\end{array}} \right).
\end{equation}
We will use this formalism in the following section for different cases to find the Casimir energy and torque.

\section{Casimir torque between two concentric cylinders}
In this section we will consider two  concentric cylinders with different types of potentials to calculate the Casimir torque.

\subsection{Semi-transparent cylinders with sinusoidal potential}
\begin{figure}[h!]
\center{\includegraphics[width=0.5\linewidth]{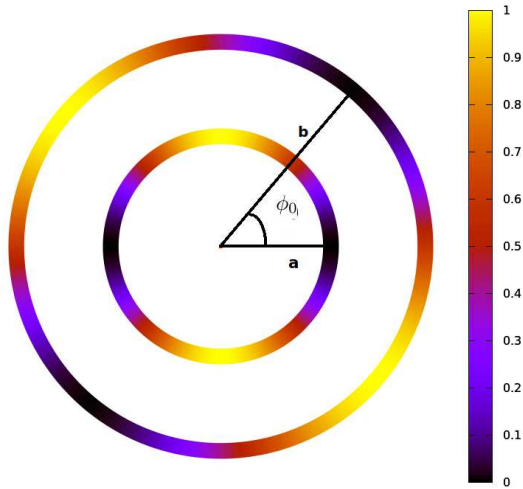}} 
\caption{Schematic geometry of two semi-transparent cylinders with sinusoidal potential and radii $a$ and $b$. The rotation angle between the cylinders is $\phi_0$. }
\label{fig1}
\end{figure}

Suppose that we have a configuration of concentric semi-transparent cylinders which is shown in Fig.~\ref{fig1}. 
Semi-transparent systems can be described by $\delta$-functions so we can write the potentials as:
\begin{equation}
{{V}_{a}}={{\lambda }_{0}}{{\sin }^{2}}(\phi) \delta (\rho-a),
\end{equation}
\begin{equation}
{{V}_{b}}={{\lambda }_{0}}{{\sin }^{2}}(\phi -{{\phi }_{0}})\delta (\rho-b),
\end{equation}
where $a$ and $b$ identify the inner and outer cylinders respectively 
and ${{\lambda }_{0}}$ is the  coupling parameter between two cylinders.
The rotation angle between the cylinders is $\phi_0$.

It is now possible to use the solution of equation  (\ref{eq5}) in a cylindrical coordinate system that can be written as a linear 
combination of ${{I}_{n}}(\kappa \rho){{e}^{in\phi }}{{e}^{i{{k}_{z}}z}}$ and  ${{K}_{n}}(\kappa \rho){{e}^{in\phi }}{{e}^{i{{k}_{z}}z}}$. 
Here ${{I}_{n}}$ and ${{K}_{n}}$  are the modified Bessel function of the first and third kind respectively  and ${{\kappa }^{2}}={{\bar{\kappa} }^{2}}+k_{z}^{2}$. 
Since the two cylinders are concentric, equations (\ref{green}) and (\ref{phi_o}) should  be equivalent; this implies the following translation matrix,
\begin{equation}
\begin{array}{l}
 X^{ab}  = \left( {\begin{array}{*{20}c}
   0 & {\hat 1}  \\
   0 & 0  \\
\end{array}} \right) \\ 
 X^{ba}  = \left( {\begin{array}{*{20}c}
   0 & 0  \\
   {\hat 1} & 0  \\
\end{array}} \right). \\ 
 \end{array}
\end{equation}
For the interior cylinder, ${{I}_{n}}$ is the regular solution and for the exterior cylinder ${{K}_{n}}$ is the outgoing solution.
Therefore the scattering amplitudes are given by:
\begin{equation}
{{(F_{a}^{ee})}_{mn}}=\left\langle \left. {{I}_{n}}(\kappa \rho){{e}^{in\phi }} \right|T\left| {{I}_{m}}(\kappa \rho){{e}^{im\phi }} \right. \right\rangle, 
\end{equation}
and
\begin{equation}
{{(F_{b}^{ii})}_{mn}}=\left\langle \left. {{K}_{n}}(\kappa \rho){{e}^{in\phi }} \right|T\left| {{K}_{m}}(\kappa \rho){{e}^{im\phi }} \right. \right\rangle.
\end{equation}
Now, we  will   follow  the problem in two different regimes.
\subsubsection{Weak coupling regime:}
In the weak coupling regime, $\lambda_0$ is small, so in the Lippmann-Schwinger expansion we only keep the first term, which leads to $T\approx V$. Using  the $\ln \left( 1-N \right)\cong -N$ approximation we can write the energy per unit length ($L=1$),
\begin{equation}
   \varepsilon =\frac{\hbar c L}{4\pi }\int_{0}^{\infty }{d\kappa \kappa \mathrm{Tr}\hspace{1pt} (N(\kappa))},
\end{equation}
in which we used $N(\kappa) = F_a^{^{ee}}(\kappa) F_b^{^{ii}}(\kappa)$.
So one can obtain the torque  from derivative of energy,
\begin{equation}
   \tau =\frac{-\partial \varepsilon }{\partial \phi_0 }=\frac{\hbar c}{4\pi }\int\limits_{0}^{\infty }{d\kappa \kappa \frac{\partial }{\partial \phi_0 } \mathrm{Tr}N}.
\end{equation}
 
In this regime, the scattering matrix elements, due to the non-vanishing couplings of Fourier modes,  are given by:
\bea 
&& {{\left( F_{a}^{ee} \right)}_{mn}}=\nonumber \\
&&{{\lambda }_{0}}\int_{0}^{\infty }{\int_{0}^{2\pi }{{{I}_{m}}(\kappa \rho){{I}_{n}}(\kappa \rho){{\sin }^{2}}(\phi) \delta(\rho-a){{e}^{im\phi }}{{e}^{-in\phi }}\rho d\rho d\phi }} \nonumber \\ 
&& \ =\pi {{\lambda }_{0}}a({{I}_{n}}^{2}(\kappa a){{\delta }_{n,m}}-{{I}_{n}}(\kappa a){{I}_{n+2}}(\kappa a){{\delta }_{n+2,m}}\nonumber \\
&&-{{I}_{n}}(\kappa a){{I}_{n-2}}(\kappa a){{\delta }_{n-2,m}}), 
\eea
and
\bea 
&&{{\left( F_{b}^{ii} \right)}_{mn}}=\nonumber \\
&&{{\lambda }_{0}}\int_{0}^{\infty }{\int_{0}^{2\pi }{{{K}_{m}}(\kappa \rho){{K}_{n}}(\kappa \rho){{\sin }^{2}}(\phi -\phi_0 )\delta (\rho-b){{e}^{im\phi_0 }}{{e}^{-in\phi }}\rho d\rho d\phi }} \nonumber \\ 
 &&=\pi {{\lambda }_{0}}b({{K}_{n}}^{2}(\kappa b){{\delta }_{n,m}}-{{K}_{n}}(\kappa b){{K}_{n+2}}(\kappa b){{e}^{2i\phi_0 }}{{\delta }_{n+2,m}} \nonumber \\
&& -{{K}_{n}}(\kappa b){{K}_{n-2}}(\kappa b){{e}^{-2i\phi_0 }}{{\delta }_{n-2,m}}). 
\label{eq29}
\eea
Thus the Casimir torque is given by:
\bea
\tau &=&\pi {{\lambda }_{0}}^{2}ab\sum\limits_{n=-\infty }^{\infty }{\int\limits_{0}^{\infty }{d\kappa \kappa }}\left( \left[ {{I}_{n}}(\kappa a){{I}_{n+2}}(\kappa a){{K}_{n}}(\kappa b){{K}_{n+2}}(\kappa b) \right. \right) \nonumber\\
&& +{{I}_{n}}(\kappa a){{I}_{n-2}}(\kappa a){{K}_{n}}(\kappa b){{K}_{n-2}}(\kappa b)])\sin (2\phi_0), 
\eea
which leads to zero torque at $\phi_0=0,\pi/2$ and maximum torque at $\phi_0=\pi/4$.
It should be noted that the equilibrium state occurs whenever the points either with both maximum energies  or  with  maximum and minimum energies are close to each other. 
In the former case we have unstable equilibrium while in the latter case the equilibrium state is stable.
Furthermore, the midpoint $\phi_0=\pi/4$ corresponds to the maximum torque which is the same as in the strong coupling regime which we will discuss below. 
\subsubsection{Strong coupling regime (exact solution):}

\begin{figure}[h!]
\center{\includegraphics[width=0.55\linewidth]{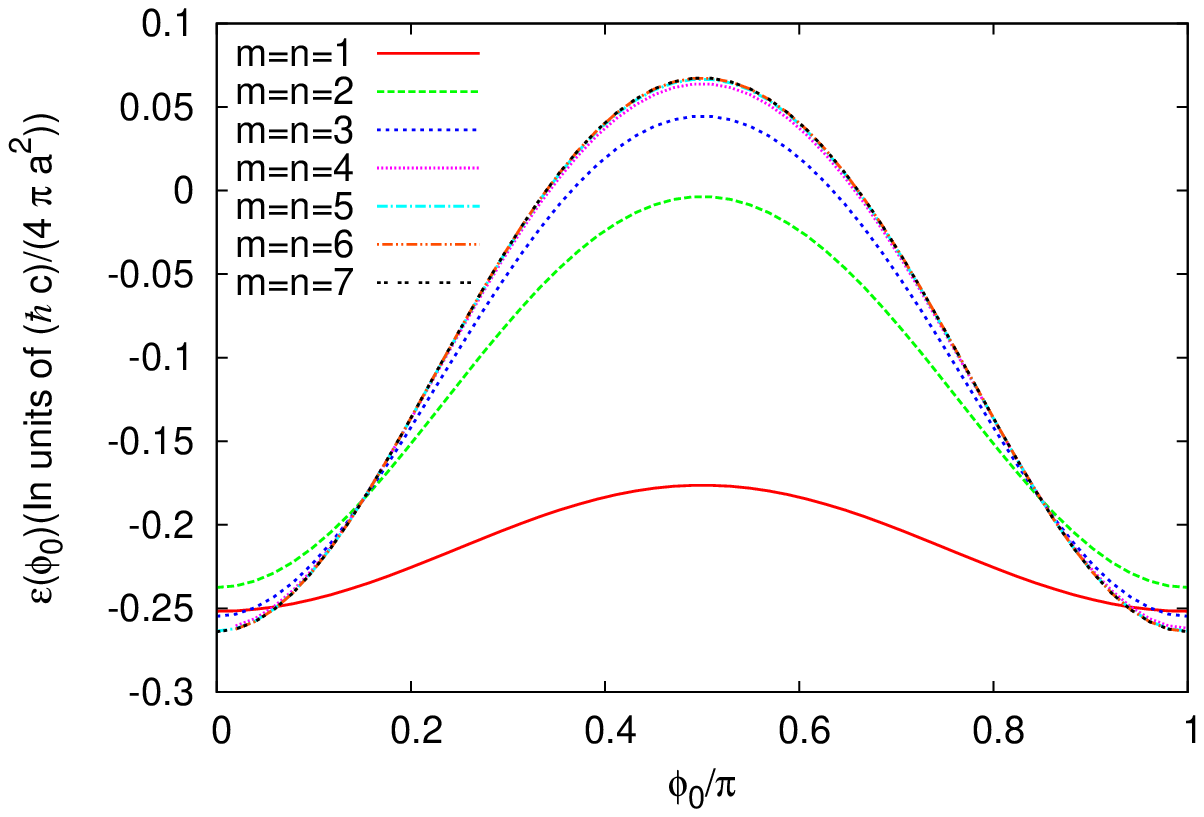}} 
\center{\includegraphics[width=0.55\linewidth]{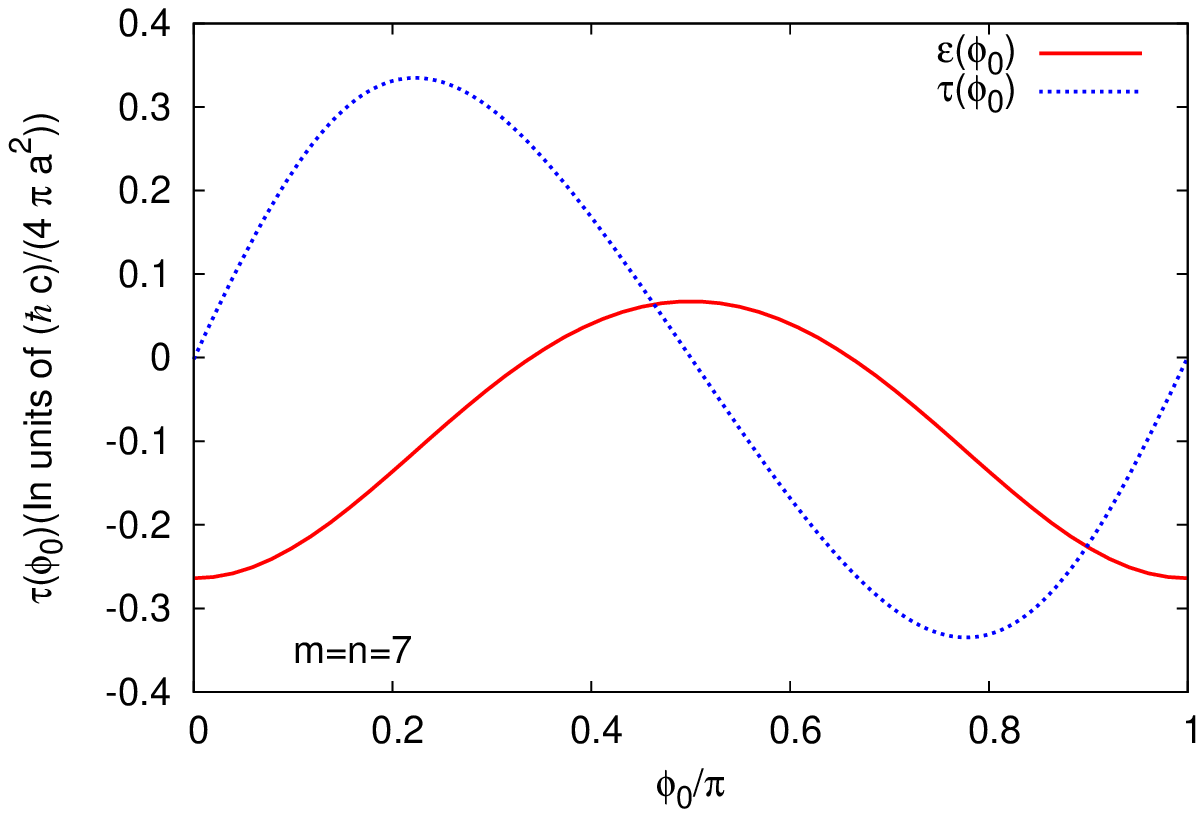}} 
\caption{(Top) Casimir energy $\varepsilon(\phi_0)$  for two concentric cylinders with sinusoidal potentials and $b=2a$ for different orders $m$ (and $n=m$) in the strong coupling regime.
(Bottom) Casimir torque and energy $\tau(\phi_0)$ with the converged $m=7$ order for the same configuration.}
\label{fig2}
\end{figure}

In the strong coupling regime, $\lambda_0\rightarrow 1$ and we have to calculate $T$ exactly.
According to the geometry of our configuration, $\rho'$ describes the point on the interior cylinder and $\rho$ corresponds to its outside space $(\rho>\rho')$, so we can  write the  solution of equation (\ref{g0}) in the cylindrical coordinate system as,
\begin{equation}
{{G}_{0}}(\kappa ,x,{x}')=\sum\limits_{n}{{{e}^{in(\phi -{\phi }')}}}{{I}_{n}}(\kappa \rho){{K}_{n}}(\kappa {\rho}'),
\end{equation}
and substitute this into equation (\ref{eq11}) which leads to the following equation. 
So, for the interior cylinder we have,
\begin{equation}
{{\left( V{{G}_{0}}V \right)}_{mn}}=\left\langle \left. {{I}_{n}}(\kappa \rho){{e}^{in\phi }} \right|V{{G}_{0}}V\left| {{I}_{m}}(\kappa \rho){{e}^{im\,\phi }} \right. \right\rangle, 
\end{equation}
which results in,
\bea
&&\quad {{\left( V{{G}_{0}}V \right)}_{mn}}={{\pi }^{2}}{{\lambda }_{0}}({{I}_{n}}(\kappa a)({{\delta }_{n,m}}-{{\delta }_{n,m+2}}-{{\delta }_{n,m-2}})\nonumber \\
&& \times {{G}_{0}}({{\delta }_{n,m}}-{{\delta }_{n,m+2}}-{{\delta }_{n,m-2}}){{I}_{m}}(\kappa a)).
\eea
  
Defining $A$ and $G_0$ matrices as
\begin{equation}
A_{nm}=({{\delta }_{n,m}}-{{\delta }_{n,m+2}}-{{\delta }_{n,m-2}}),
\end{equation}
and
\begin{equation}
{{\left( {{G}_{0}}(x) \right)}_{nm}}={{I}_{n}}(\kappa x){{K}_{n}}(\kappa x)\,{{\delta }_{n,m}}\quad \quad \quad \quad x=a,b,
\end{equation}
we can simplify the scattering amplitude
\begin{equation}
\hat{F}_{a}^{ee}=\hat{I}(\kappa a)\hat{A}(\hat{1}+{{{\hat{G}}}_{0}}(x)\hat{A})^{-1}\hat{I}(\kappa a)
\label{eq36}
\end{equation}  
in which 
\begin{equation}
{{\left( I(\kappa a) \right)}_{nm}}={{I}_{n}}(\kappa a)\ {{\delta }_{n,m}}.
\end{equation}
Similarly, for the exterior cylinder we have
\begin{equation}
{{\left( V{{G}_{0}}V \right)}_{mn}}
={{\pi }^{2}}{{\lambda }_{0}}({{K}_{n}}(\kappa b)B\,{{{\hat{G}}}_{0}}B{{K}_{m}}(\kappa b)), 
\end{equation}
and
\begin{equation}
\hat{F}_{b}^{ii}=\hat{K}(\kappa b)\hat{B}(\hat{1}+{{{\hat{G}}}_{0}}(x)\hat{B})^{-1}\hat{K}(\kappa b),
\label{eq39}
\end{equation}
where
\begin{equation}
B_{nm}=({{\delta }_{n,m}}-{{\delta }_{n,m+2}}{{e}^{2i\phi_0 }}-{{\delta }_{n,m-2}}{{e}^{-2i\phi_0 }}),
\end{equation}
and
\begin{equation}
{{\left( \hat{K}(\kappa b) \right)}_{nm}}={{K}_{n}}(\kappa b)\ {{\delta }_{n,m}}.
\end{equation}
Now we can substitute equations (\ref{eq36}) and (\ref{eq39}) into (\ref{eq4}) and take the derivative with respect to variable $\phi_0$ to obtain the Casimir torque numerically. Computing this expression for different orders of $m$ (and $n=m$), we find that for $m>4$ we will converge to the stable solution in which we have only negligibly small  changes in energy and torque. These results are shown in Fig.~\ref{fig2}.
It is clear from the plots that we have the same equilibrium points as we had in the weak coupling regime.  

\subsection{Semi-transparent cylinders with step function potential}
\begin{figure}[h]
\center{\includegraphics[width=.4\linewidth]{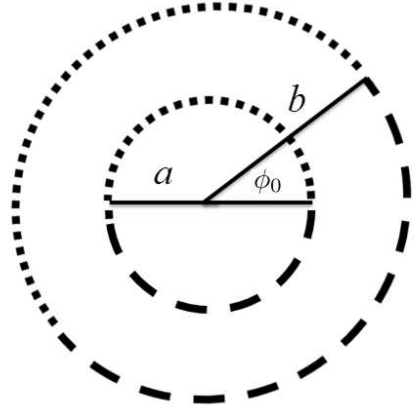}} 
\caption{Schematic geometry of two semi-transparent cylinders with step function potential and radii $a$ and $b$.}
\label{fig3}
\end{figure}

The next configuration that we will consider is shown in Fig.~\ref{fig3}.
The interior cylinder  is described by potential
\begin{equation}
V_a  = \left\{ \begin{array}{l}
 \lambda _0 \delta (\rho - a)\;\;\;\;\;\;\;\;\;\;\;\;\;\;\;\;\;\;0 \le \phi  \le \pi  \\ 
 \alpha \lambda _0 \delta (\rho - a)\;\;\;\;\;\;\;\;\;\;\;\;\;\;\;\;\pi  \le \phi  \le 2\pi \;\; \\ 
 \end{array} \right.,
\label{eq42}
\end{equation}
and for the exterior cylinder we have
\begin{equation}
V_b  =\left\{ \begin{array}{l}
 \lambda _0 \delta (\rho - b)\;\;\;\;\;\;\;\;\;\;\;\;\;\;\;\;\;\;\;\;\phi_0  \le \phi  \le \phi_0  + \pi  \\ 
 \alpha \lambda _0 \delta (\rho - b)\;\;\;\;\;\;\;\;\;\;\;\;\;\;\;\;\;\;\phi_0  + \pi  \le \phi  \le \phi_0  + 2\pi. \;\; \\ 
 \end{array} \right.
\label{eq43}
\end{equation}

\subsubsection{Weak coupling regime:}

\begin{figure}[h]
\center{\includegraphics[width=0.6\linewidth]{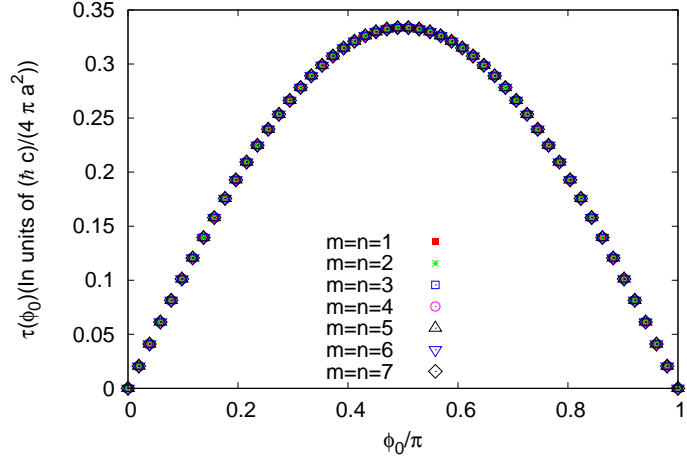}} 
\caption{Casimir torque,  $\tau(\phi_0)$, for two concentric cylinders with step function potentials for different orders $m$ (and $n=m$) in the weak coupling regime.}
\label{fig4}
\end{figure}

As  discussed above, we can consider two different regimes.
In the weak coupling regime, the scattering amplitudes are given by:
\begin{equation}
 \left( {F_a^{ee} } \right)_{mn}  =\left\{ \begin{array}{l}
 \frac{{ - 2a\lambda _0 }}{{i(m - n)}}I_n (\kappa a)I_m (\kappa a)(1 - \alpha )\;\;\;\;\;m - n = 2l + 1 \\ 
 \pi a\lambda _0 I_n ^2 (\kappa a)(1 + \alpha )\;\;\;\;\;\;\;\;\;\;\;\;\;\;\;\;\;\;\;m = n \\ 
0\;\;\;\;\;\;\;\;\;\;\;\;\;\;\;\;\;\;\;\;\;\;\;\;\;\;\;\;\;\;\;\;\;\;\;\;\;\;\;\;\;\;\;\;\;\;\;m - n = 2l \\ 
 \end{array} \right.
\end{equation}
for $a$ cylinder and 
\begin{equation}
\left( {F_b^{ii} } \right)_{mn}  ={\kern 1pt} {\kern 1pt} \left\{ \begin{array}{l}
 \frac{{ - 2\lambda _0 be^{i(m - n)\theta } }}{{i(m - n)}}K_n (\kappa b)K_m (\kappa b)(1 - \alpha )\;;m - n = 2l + 1 \\ 
\pi \lambda _0 bK_n ^2 (\kappa b)(1 + \alpha )\;\;\;\;\;\;\;\;\;\;\;\;\;\;\;\;\;\;\;\;\;\;\;\;\;\;;\;m = n \\ 
 0\;\;\;\;\;\;\;\;\;\;\;\;\;\;\;\;\;\;\;\;\;\;\;\;\;\;\;\;\;\;\;\;\;\;\;\;\;\;\;\;\;\;\;\;\;\;\;\;\;\;\;\;\;\;;m - n = 2l \\ 
 \end{array} \right.
\end{equation}
for $b$ cylinder where $l$ is an integer.


\begin{figure}[h]
\center{\includegraphics[width=0.5\linewidth]{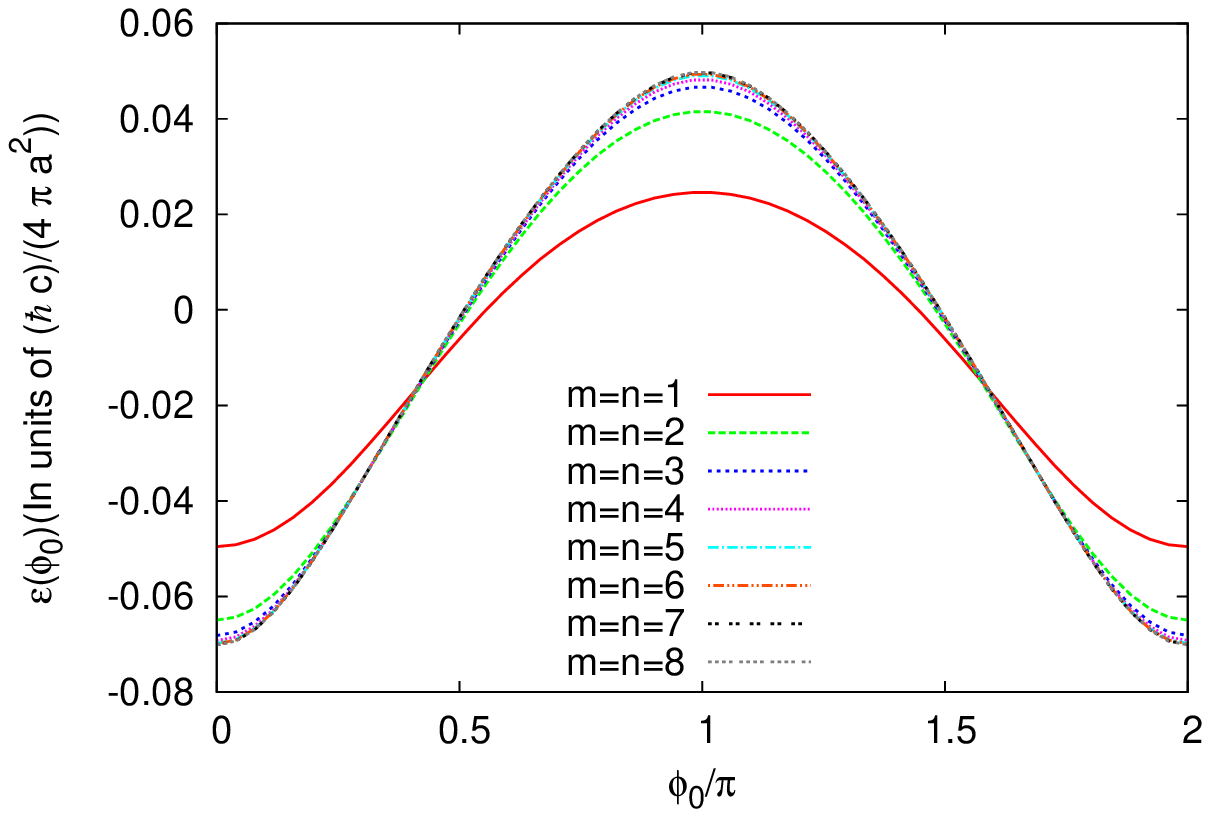}} 
\center{\includegraphics[width=0.5\linewidth]{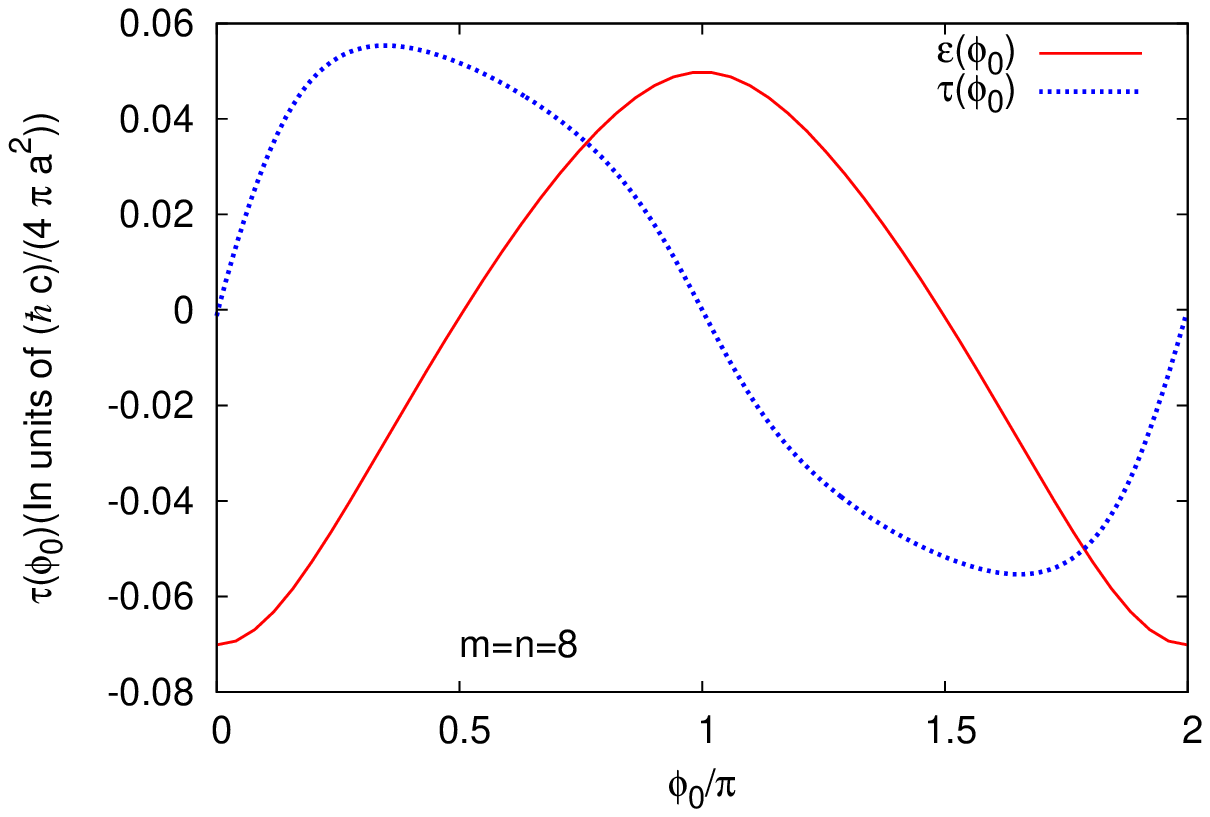}} 
\caption{(Top) Casimir energy $\varepsilon(\phi_0)$  for two concentric cylinders with step function potentials for different orders $m$ (and $n=m$) in the strong coupling regime.
(Bottom) Casimir torque and energy $\tau(\phi_0)$ with the converged $m=n=8$ order for the same configuration.}
\label{fig5}
\end{figure}

Therefore one can compute the Casimir torque as
\bea
&&\tau  = \frac{{\lambda _0 ^2 (1 - \alpha )^2 ab}}{{\pi (m - n)}}  \\
&&\times \sum\limits_{m =  - \infty }^\infty  {\sum\limits_{n =  - \infty }^\infty  {\int\limits_0^\infty  {d\kappa {\kern 1pt} \kappa \left( {I_n (\kappa a)I_m (\kappa a)K_n (\kappa b)K_m (\kappa b)} \right)} } } \sin ((m - n)\phi_0), \nonumber
\eea
which results to the plots of Fig.~\ref{fig4} for different orders $m$ (and $n=m$) versus $\phi_0/\pi$.
It shows that in the weak coupling regime, the torque is approximately constant for all $m$ orders and  reaches its equilibrium state  at angles $\phi_0=0,\pi,2\pi$ so that for $\pi< \phi_0 < 2\pi$ a negative Casimir torque tends to reduce the angle. 

\subsubsection{Strong coupling regime(exact solution):}

In the strong coupling regime, we can do the same calculation as before and the only difference is in the expression we have for $A$ and $B$ matrices,
\bea
A_{n,m} &=& \delta _{n,m}  + \mathop \sum \limits_{c = 0}^\infty  \left[ {\frac{1}{{\left( {2c + 1} \right){\rm i}}}\delta _{n + 2c + 1,m} } \right] \nonumber \\ 
&&+ \mathop \sum \limits_{c = 0}^\infty  \left[ {\frac{{ - 1}}{{\left( {2c + 1} \right){\rm i}}}\delta _{n - (2c + 1),m} } \right],
\eea
\bea
B_{n,m} &=& \delta _{n,m}  + \mathop \sum \limits_{c = 0}^\infty  \left[ {\frac{{{\rm e}^{\left( {2c + 1} \right){\rm i}\phi_0 } }}{{\left( {2c + 1} \right){\rm i}}}\delta _{n + 2c + 1,m} } \right] \nonumber \\
&&+ \mathop \sum \limits_{c = 0}^\infty  \left[ {\frac{{ - {\rm e}^{ - \left( {2c + 1} \right){\rm i}\phi_0 } }}{{\left( {2c + 1} \right){\rm i}}}\delta _{n - (2c + 1),m} } \right],
\eea
which can be computed numerically if one substitutes it in equation (\ref{eq4}).

Fig.~\ref{fig5} shows the resulting plots of Casimir energy for different orders $m$ (and $n=m$). It also shows the Casimir torque corresponds to the converged solution. It is obvious that the torque is zero for angles $\phi_0=0,\pi,2\pi$ again which are the equilibrium states of the cylinders. Fig.~\ref{fig5} also shows that the torque reaches the maximum for $\phi_0\approx \pi/2$.

Before ending this section, we make some remarks about the effect of different terms on the power series of the solutions.
It is clear that for various potentials in the weak coupling regime, adding higher-order terms have no effect on the solution while in the strong coupling regime, we observe that the solutions are order dependent. In order to understand this, we can say that, in the strong coupling limit, the expression of equation (\ref{eq39}) (compared to equation (\ref{eq29}) in the weak coupling regime) contains an order-dependent  denominator which affects the solutions.   

\begin{figure}[h]
\center{\includegraphics[width=.4\linewidth]{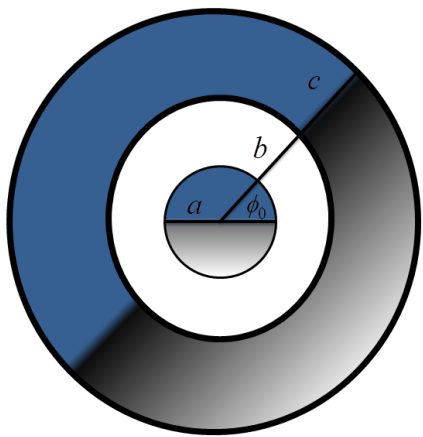}} 
\caption{Schematic geometry of two concentric dielectric  cylinders,   one with  $0<\rho<a$ and other with $b<\rho<c$.}
\label{fig6}
\end{figure}

\subsection{Dielectric cylinders}
\begin{figure}[h]
\center{\includegraphics[width=.6\linewidth]{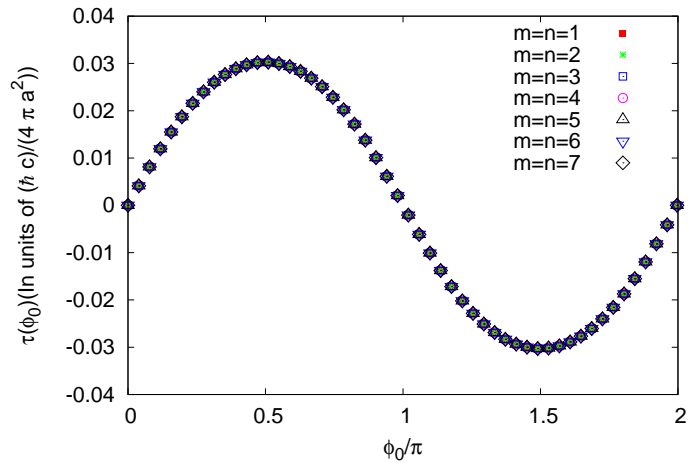}} 
\caption{Casimir torque,  $\tau(\phi_0)$, for two concentric  dielectric cylinders  for different orders $m$ (and $n=m$) at the weak coupling regime.}
\label{fig7}
\end{figure}

Finally, we will consider the case of two dielectric cylinders, which is shown in Fig.~\ref{fig6} and  can be described by the following potentials:
\begin{equation}
V_a  = \left\{ \begin{array}{l}
\lambda _0 \Theta (\rho - a)\;\;\;\;\;\;\;\;\;\;\;\;\;\;\;\;\;\quad \quad \quad \quad \quad 0 \le \phi  \le \pi  \\ 
\alpha \lambda _0 \Theta (\rho - a)\;\;\;\;\;\;\;\;\;\;\;\;\;\;\;\quad \quad \quad \quad \quad \pi  \le \phi  \le 2\pi \;\; \\ 
 \end{array} \right.,
\end{equation}
\begin{equation}
 V_b  = \left\{ \begin{array}{l}
\lambda _0 \left( {\Theta (\rho - b) + \Theta (c - \rho)} \right)\;\;\;\;\;\;\;\;\;\phi_0  \le \phi  \le \phi_0  + \pi  \\ 
 \alpha \lambda _0 \left( {\Theta (\rho - b) + \Theta (c - \rho)} \right)\;\;\;\;\;\;\;\phi_0  + \pi  \le \phi  \le \phi_0  + 2\pi. \;\; \\ 
 \end{array} \right.
\end{equation}
In this case we only consider the weak coupling regime. 
In this regime, we can write the scattering amplitudes for $a$ cylinder as,
\begin{equation}
\left( {F_a^{ee} } \right)_{mn}={\kern 1pt} {\kern 1pt} \left\{ \begin{array}{l}
  \frac{{ - 2a\lambda _0 (1 - \alpha )}}{{i(m - n)}}\int\limits_0^a {I_n (\kappa r)I_m (\kappa \rho)} \rho d\rho\;\;\;\;\;m - n = 2l + 1 \\ 
 \pi a\lambda _0 (1 + \alpha )\int\limits_0^a {I_n (\kappa r)I_m (\kappa \rho)} \rho d\rho\;\;\;\;\;\;\;m = n \\ 
 \;\;0\;\;\;\;\;\;\;\;\;\;\;\;\;\;\;\;\;\;\;\;\;\;\;\;\;\;\;\;\;\;\;\;\;\;\;\;\;\;\;\;\;\;\;\;\;\;\;\;m - n = 2l \\ 
 \end{array} \right.,
\end{equation}
and for $b$ cylinder
\begin{equation}
\left( {F_b^{ii} } \right)_{mn}  =\left\{ \begin{array}{l}
\frac{{ - 2\lambda _0 be^{i(m - n)\phi_0 } (1 - \alpha )}}{{i(m - n)}}\int\limits_b^c {K_n (\kappa \rho)K_m (\kappa \rho)} \rho d\rho\;\;\;\;\;\;\;\;,m - n = 2l + 1 \\ 
= \pi \lambda _0 b(1 + \alpha )\int\limits_b^c {K_n (\kappa \rho)K_m (\kappa \rho)} \rho d\rho\;\;\;\;\;\;\;\;\;\;\;\;\quad m = n \\ 
 0\;\;\;\;\;\;\;\;\;\;\;\;\;\;\;\;\;\;\;\;\;\;\;\;\;\;\;\;\;\;\;\;\;\;\;\;\;\;\;\;\;\;\;\;\;\;\;\;\;\;\;\;\;\;\;\;\;\;\;\;\;\;\;\;\quad \quad m - n = 2l \\ 
 \end{array} \right. .
\end{equation}
This results in the following Casimir torque,
\be
\tau  = \frac{{\lambda _0 ^2 (1 - \alpha )^2 ab}}{{\pi (m - n)}} \sum\limits_{m =  - \infty }^\infty  {\sum\limits_{n =  - \infty }^\infty  {\int\limits_b^c {\int\limits_0^a {\int\limits_0^\infty 
{d\kappa \;d\rho\;dh{\kern 1pt} \kappa \,\rho\,h\left( {I_n (\kappa \rho)I_m (\kappa \rho)K_n (\kappa h)K_m (\kappa h)} \right)} } } } } \sin (m - n)\phi_0, 
\ee
which can be computed numerically.
Fig.~\ref{fig7}  shows the plots of Casimir torque for different orders $m$ (and $n=m$) versus $\phi_0/\pi$.
Clearly the torque in the weak coupling regime is independent of the order $m$ with zero value for $\phi_0 =0,\pi,2\pi$.

\section{Conclusion} 
In this paper, we have employed the multiple scattering approach to study the 
Casimir torque for some examples of  two inhomogeneous concentric cylinders.  
For simplicity we have considered a massless scalar field for three different potentials to find the Casimir energy and torque both analytically and numerically in the weak and strong coupling regimes. 
For all cases, we have computed the $\tau(\phi_0)$,   which shows that the resulting torque is zero for $\phi_0=0,\pi$ and the maximum torque   is attained when the rotating angle $\phi_0$ is exactly or approximately equal to $\pi/2$ depending on coupling regimes. 
The equilibrium state of the system takes place at the same point in weak and strong coupling regimes for both sinusoidal and step function potentials while the position of maximum torque  in the former configuration is the same for both regimes but for the latter configuration we have observed some small changes.  

Finally, our findings suggest a new scenario in designing nano rotors, to use  inhomogeneous surfaces instead of corrugated one which seems to be easier in practice. The extension of this model for finite temperature and the dynamic Casimir
effect ~\cite{Sar1,Sar2} would also be interesting.


\end{document}